\documentclass[aps,prl,preprint,groupedaddress]{revtex4}

\usepackage{graphicx}
\usepackage{dcolumn}
\usepackage{bm}

\begin{document}


\title{$\alpha$- and $\beta$- Relaxation Dynamics of a fragile plastic crystal}



\author{L. C. Pardo}
\affiliation{Experimental Physics V, Center for Electronic
Correlations and Magnetism, University of Augsburg, 86135
Augsburg, Germany and Dept. F\'{\i}sica i Enginyeria Nuclear,
E.T.S.E.I.B., Universitat Polit\`ecnica de Catalunya, Diagonal
647, 08028 Barcelona, Catalonia, Spain.}
\author{P. Lunkenheimer and A. Loidl}
\affiliation{ Experimental Physics V, Center for Electronic
Correlations and Magnetism, University of Augsburg, 86135
Augsburg, Germany}

\date{\today}

\begin{abstract}

We present a thorough dielectric investigation of the relaxation
dynamics of plastic crystalline Freon112, which exhibits freezing
of the orientational degrees of freedom into a glassy crystal
below 90 K. Among other plastic crystals, Freon112 stands out by
being relatively fragile within Angell's classification scheme and
by showing an unusually strong  $\beta$-relaxation. Comparing the
results to those on Freon112a, having only a single molecular
conformation, points to the importance of the presence of two
molecular conformations in Freon112 for the explanation of its
unusual properties.
\end{abstract}


\maketitle

\section{Introduction}
Glass phenomena occur when a dynamically disordered system (e. g.
aliquid, plastic crystal or paramagnet) freezes as a function of
external temperature or pressure devoid of long range order. In
this freezing process, one or more degrees of freedom of atoms or
molecules continuously slow down, reaching the so called glass
transition when their dynamics has a characteristic time,
generally chosen to be $10^{2}\textrm{ s}$ (for recent reviews on
glass transition see, e.g., \cite{glass}). Since in the liquid
phase there is translational and orientational disorder, the glass
transition of canonical glass formers is associated with the
freezing of these two degrees of freedom. But a mesophase can
exist between the completely ordered crystalline phase and the
translationally and orientationally disordered liquid phase, the
so called plastic phase or orientationally disordered (OD) phase
\cite{parsonage,sherwood}. In the plastic phase, the centers of
mass of the molecules have spatial long range order, forming a
lattice which generally has high symmetry (such as cubic,
quasi-cubic or rhombohedral \cite{parsonage,sherwood,cloros}), but
there is only short-range order with respect to the orientational
degrees of freedom \cite{cl1cl4}. As for glass-forming liquids,
the typical phenomenology of a glass transition also can be
realized in various plastic crystals when the temperature is
decreased, but in this case only the orientational degrees of
freedom are frozen, yielding the formation of a "glassy crystal"
\cite{adachi1968}, also called OD glass. \\

Concerning the dynamics of OD phases, dielectric spectroscopy has
been revealed to be a useful tool to understand the complex
dynamics of these phases
\cite{PCdynamics,leslie1994,c8,puertas,brand1999,miller1998,amoureux1984}.
Based on these works there seem to be some general dynamic
features of OD phases and its glasses: They are rather strong
(following the definition of Angell \cite{angell1985}) and they
follow the B\"{o}hmer relation between non-exponentiality of the
$\alpha$ relaxation and fragility
\cite{bohmerrelation,bohmer1999}. On the contrary, the so called
Nagel scaling \cite{nagelscaling} does not seem to work for these
phases \cite{schneider2000,brand1999,PCdynamics}. In addition
there is only a weak or no $\beta$-relaxation at all and the
excess wing, showing up as a second power law at the high
frequency flank of the $\alpha$ peak in many canonical glass
formers
\cite{nagelscaling,lunkenheimer2000,hofmann1994,kudlik1999}, is
either absent in plastic crystals \cite{brand1999} or can be
ascribed to a weak secondary relaxation \cite{PCdynamics,c8}.\\

Among compounds forming orientational glasses,
1,2-difluoro-1,1,2,2-tetrachloroethane ($CFCl_{2}-CFCl_{2}$), also
named Freon112, has been revealed to be an exception to the
aforementioned strongness of glassy crystals \cite{freonstrong}.
When cooled from the liquid phase, a BCC OD phase is formed
\cite{kishimoto1978} which on further cooling yields a glassy
crystal. The completely ordered phase, whose symmetry is not
known, appears only after long waiting times (about 50 days at 77
K - 160 K) \cite{kishimoto1978}. Freon112 has two energetically
non-equivalent molecular conformations, namely trans (with a
$C_{2h}$ symmetry) and gauche (with a $C_{2}$ symmetry)
\cite{Iwasaki1957}. The trans conformer is more stable than
gauche, and while the latter has a dipolar moment of 0.26D
\cite{kagarise1955} the trans conformer is non-polar. The energy
barrier ($\Delta H^{*}$) and energy difference between the two
conformers ($\Delta H$) was determined using NMR, Raman and far
infrared spectroscopy, and specific heat measurements
\cite{newmark1965,pethrick1971,kishimoto1978} yielding values of
$\Delta H^{*}$ between 0.3 eV and 0.42 eV and $\Delta
H=0.005-0.008$ eV. In the specific heat measurements by Kishimoto
et al. \cite{kishimoto1978} a strong jump at 90 K showed up, which
was assigned to the primary glass transition. Two additional
thermal effects were observed at 60 K and 130 K, ascribed to a
secondary relaxation and the freezing of the trans-gauche
transition, respectively. The primary glass transition temperature
was also determined by Kr\"{u}ger et al. \cite{kruger1994} taking
into account the change in the slope of the temperature dependence
of the refractive index and BCC lattice parameter leading to
$T_{g}=86\textrm{ K}$. Also dielectric data are reported in the
latter work, although actual spectra are not shown and only the
temperature dependent curve of dielectric loss for a single
frequency is given. Recently a molecular dynamics (MD) simulation
has been performed on this compound \cite{affouard2005}, in which
using a simplified model of Freon112, it was possible to obtain
information on the "slow" dynamics of this compound (slow when
compared to the time domain usually investigated in MD
simulations) and therefore the simulation can be tested by
dielectric spectroscopy results. In that work Mode Coupling Theory
(MCT) \cite{gotze1992} was also successfully employed to interpret the results.\\

In the present work also the Freon compound
1,1-difluoro-1,2,2,2-tetrachloroethane ($CCl_{3}-CClF_{2}$, also
called Freon112a) is investigated, which has the same overall
molecular formula as Freon112, but with both fluor atoms connected
to one carbon. There is only one conformation of this molecule.
Therefore, since it has a permanent dipolar moment, and it has no
trans-gauche disorder, it is of interest to compare its properties
with those of the Freon112 molecule. No information, neither on
the relaxation dynamics nor on the polymorphism of this compound,
is found in the bibliography.

\section{Experimental details}
Freon112 and Freon112a were purchased from ACBR company with a
minimum purity of 99\%, and were used without further
purification. Dielectric spectra were obtained by means of
frequency response analysis using a Novocontrol-$\alpha$ analyzer
in the frequency range 10 mHz $\leq \nu \leq$ 1 MHz , and by means
of a reflectometric technique using an HP4291 impedance analyzer
at 1 MHz $\leq \nu \leq 1.8$ GHz \cite{schneider2001}. Cooling and
heating of the samples were performed using a nitrogen gas-heating
system and a helium-based cryostat. The parallel-plate capacitors
were filled in the liquid phase and, because of the high vapor
pressure of these compounds,
quickly cooled down in order to avoid sublimation.\\
Specific heat measurements were done by means of an AC-technique
using the PPMS from Quantum Design. The  cell was filled with
about 30 mg of the substance in the liquid phase. Contributions
from the specific heat of the cell were subtracted from that
experimentally determined. Additional calorimetric experiments
were performed by a Differential Scanning Calorimetry (DSC) using
a Perkin Elmer DSC7 device with high pressure cells, because of
the high vapor pressure of the substances.

\section{Results}
\subsection{Freon112}

\begin{figure}
\includegraphics[width=0.75\textwidth]{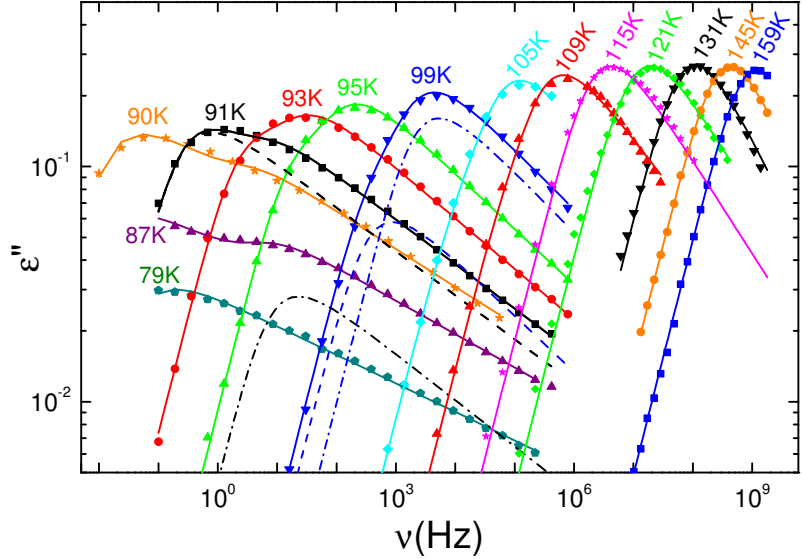}
\caption{\label{efreon12} \footnotesize Dielectric loss for the
compound Freon 112; lines show the fits using two ($79 \textrm{ K}
< T< 130 \textrm{ K}$) or one ($130 \textrm{ K} < T < 160 \textrm{
K}$) CD functions. The two CD functions used to fit the spectra in
the first temperature range are indicated for 91 K and 99 K by the
dashed and dash-dotted lines.}
\end{figure}

\begin{figure}
\includegraphics[width=0.75\textwidth]{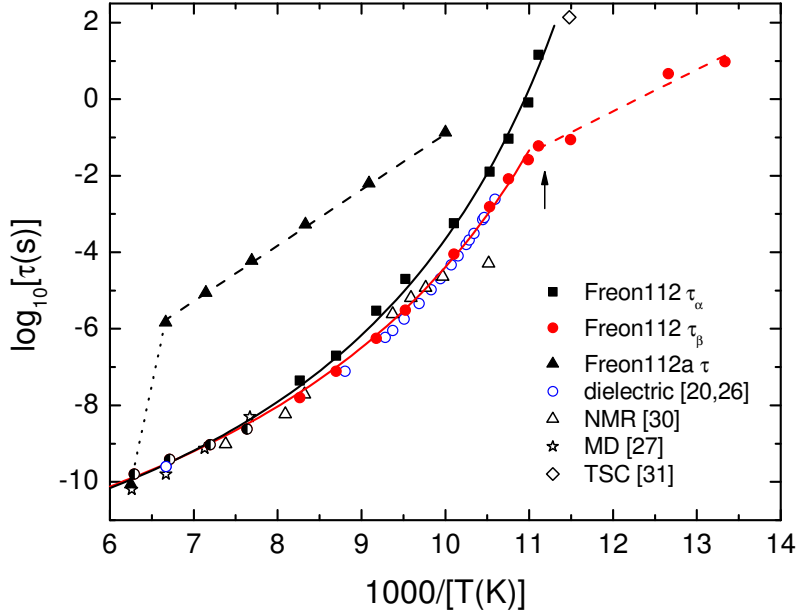}
\caption{\label{logtaug} \footnotesize Relaxation time for
$\alpha$, $\beta$ and $\alpha\beta$ (half filled circles) peaks in
Freon112 and $\alpha$ peak for Freon112a. Dielectric spectroscopy
data of Kr\"{u}ger et al. \cite{kruger1994} and Angell
\cite{freonstrong}, NMR data \cite{stokes79}, TSC data
\cite{fuchs1985} and MD data \cite{affouard2005} are also shown.
The solid and dashed lines are fits with the VFT law, eq.(1), and
an Arrhenius behavior, respectively. The arrow indicates the glass
temperature.}
\end{figure}

The main feature showing up in the spectra is a well-pronounced
relaxation peak (Fig. \ref{efreon12}) \cite{spectra}. Its
continuous shift over many decades towards lower frequencies with
decreasing temperature, mirrors the glassy freezing of the
orientational dynamics in this plastic crystal \cite{PCdynamics}.
At higher temperatures, $T >$ 100 K, a single peak is observed,
which can be well described, e.g., by the empirical Cole-Davidson
(CD) function (lines in Fig. \ref{efreon12}). However, for $T <$
100 K a shoulder emerges at the low-frequency flank of the
relaxation peak, which seems to increase strongly in intensity
and, with further lowering of temperature, develops into a
separate peak with an amplitude even larger than that of the
high-frequency peak. Tentatively we will denote the slow process
as $\alpha$-relaxation and the faster process as
$\beta$-relaxation; at high temperatures, where both processes are
merged, the term $\alpha\beta$-process is chosen. Overall, while
in earlier works only a single relaxation process was reported
\cite{freonstrong,kruger1994,affouard2005,stokes79,fuchs1985},
obviously two processes appear in this compound, exchanging the
role of the dominating one depending on temperature. In order to
analyze the spectra, a single Cole-Davidson (CD) function was
sufficient to fit the relaxation peaks for temperatures above 130
K. But it was impossible, neither with a single CD nor with a
Havriliak-Negami (HN) function, to fit the spectra for $T < 130
\textrm{ K}$. Instead for this temperature range, even for the
curves between 105 K and 121 K where the appearance of a
low-frequency shoulder is not immediately evident, two CD
functions had to be used to fit the spectra (for 91 K and 99 K the
two constituents of these fits are indicated by the dashed and
dash-dotted lines). This finding provides further evidence for the
existence of two separate relaxation phenomena in Freon112. Now
the question arises why only a single relaxation phenomenon was
reported in previous works
\cite{freonstrong,affouard2005,fuchs1985,stokes79,kruger1994}.
Figure \ref{logtaug} shows the relaxation map including the
results of the fits of Fig. \ref{efreon12} and those reported in
the literature,
\cite{freonstrong,kruger1994,affouard2005,stokes79,fuchs1985}.
Obviously, all published data have been collected at temperatures
too high to clearly resolve the second relaxation. Only in
\cite{kruger1994}, values of $\tau$ obtained from dielectric
measurements were reported extending down to 95 K, where a small
low-frequency shoulder of the loss peak should reveal the presence
of a second relaxation process (Fig. \ref{efreon12}). However,
unfortunately in that work the data have not been analyzed in
terms of frequency-dependent plots and the relaxation times were
evaluated from the temperature-dependent data instead. As revealed
by Fig. 2, the relaxation times obtained by Kr\"{u}ger et al.
\cite{kruger1994} correspond to the  $\beta$-relaxation,
representing the dominant relaxation process in the temperature
range investigated in this work.\\

The previously determined non-Arrhenius behavior of $\tau(T)$ for
this compound is evident from the strong curvature in the
Arrhenius representation of Fig. \ref{logtaug}. In order to fit
the relaxation time as a function of temperature, a
Vogel-Fulcher-Tammann (VFT) equation was used (solid line),

\begin{equation} \label{VFT}
\tau=\tau_{0}\:exp\left[\frac{DT_{VF}}{T-T_{VF}}\right]
\end{equation}

with $T_{VF}$ the Vogel-Fulcher temperature and \textit{D} the
strength parameter \cite{angell1985}, obtaining values of
$\tau_{0}=6.2\times 10^{-14}\textrm{ s}$, $T_{VF}=68.8 \textrm{
K}$ and \textit{D} = 10. The relatively small value of the
strength parameter already indicates the fragile character of this
OD glass, which will be treated in more detail in the discussion
section. In order to determine the glass temperature we have
adopted the general definition, \emph{i.e.} the temperature for
which the relaxation time is 100 s, leading to $T_{g} = 88
\textrm{ K}$, in reasonable agreement with the previous
calorimetric determinations
 ($T_{g} = 90 \textrm{ K}$) \cite{kishimoto1978,suga1985}.
Concerning the $\beta$ relaxation process, it also follows a VFT
behavior for the temperature range above $T_{g}$
($\tau_{0}=7.2\times 10^{-14} \textrm{ s}$, $T_{VF}=64.9 \textrm{
K}$ and \textit{D} = 10.9). For lower temperatures
$\tau_{\beta}(T)$ shows a transition into a much weaker
Arrhenius-like temperature dependence. The "glass temperature"
obtained for this second relaxation is about 70 K. In this
context, it is interesting that in the specific heat measurements
of Freon112 \cite{kishimoto1978,suga1985} a "small heat-capacity
anomaly at around 60 K has been attributed to the $\beta$
relaxation". Since the determination of this temperature effect is
subjected to some error, and the slope of
$\tau_{\beta}\textrm{(T)}$ in Fig. \ref{logtaug} depends on
thermal history in this sub-$T_{g}$ region, it seems reasonable
that the aforementioned thermal effect could be related to the
$\beta$ relaxation observed by dielectric spectroscopy. The
agreement of the glass temperature determined from the
$\alpha$-relaxation time with that from specific heat measurements
and the finding of a transition of $\tau_{\beta}(T)$ into
Arrhenius behavior already at $\tau_{\beta}$-values much lower
than 100 s, corroborates the correct assignment of both
relaxations as  $\alpha$- and $\beta$-relaxation. Obviously,
despite the relaxation occurring at higher frequencies is the
dominating one at high temperatures, it is the relaxation denoted
as  $\alpha$-relaxation that determines the glassy freezing in
Freon112.\\

\begin{figure}
\includegraphics[width=0.75\textwidth]{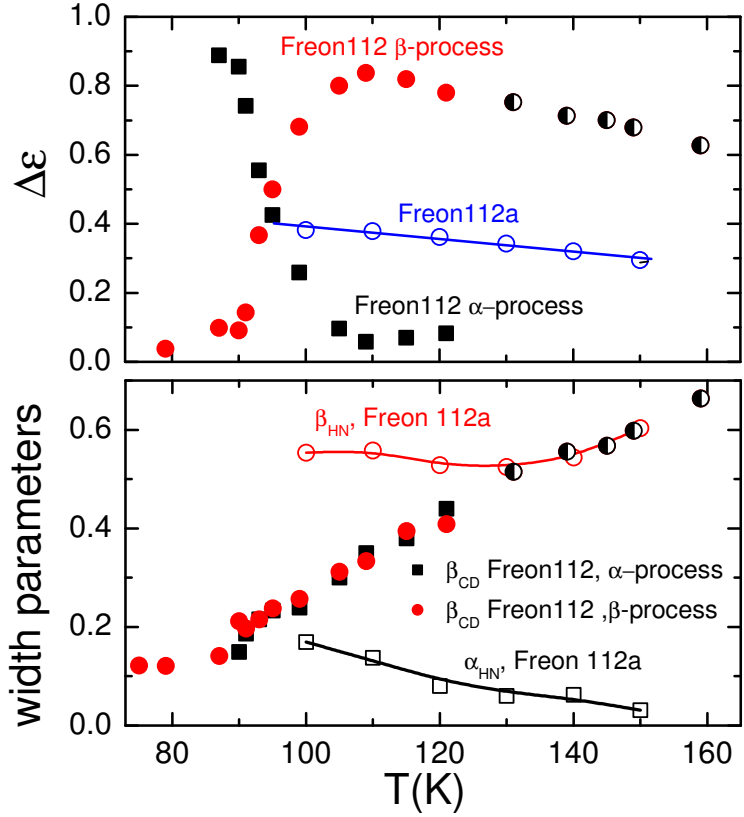}
\caption{\label{parameter} \footnotesize Relaxation strength and
width parameters used to fit Freon112 and Freon112a spectra. The
parameters used to fit the single high temperature $\alpha\beta$
relaxation peak for Freon112 are shown as half-filled circles. The
lines are drawn to guide the eyes.}
\end{figure}

In Figure \ref{parameter} the remaining CD fitting parameters for
Freon112 are shown. The results concerning the dielectric strength
($\Delta\varepsilon$) are especially interesting. When increasing
temperature the dielectric strength of the $\alpha$ relaxation
decreases quickly at about the glass transition temperature and
remains constant for higher temperatures. Because in this
temperature range the $\alpha$ relaxation only appears as a
shoulder in $\varepsilon''(\nu)$, its dielectric strength could
not be unambiguously determined but the general trend should be
correct. On the contrary $\Delta\varepsilon _{\beta}$ increases at
the same temperature, that is $T_{g}$. For temperatures above 130
K, only a single peak has been fitted, and therefore for this
temperature range only $\Delta\varepsilon _{\alpha\beta}$ could be
determined. With respect to the exponent of the high-frequency
flank of the relaxation peaks, for both $\alpha$ and
$\beta$-relaxation this parameter is roughly the same, and becomes
extremely small for low temperatures. When considering these
results, however, one should be aware that especially at high
temperatures the determination of $\beta_{CD}$ of the
$\alpha$-process  has a high uncertainty due to the strong overlap
with the  $\beta$-process. In addition, one should mention that it
is not so clear if a simple \emph{additive} superposition of
different contributions to $\varepsilon''(\nu)$ is really
justified \cite{fitting,arbe1996}. While alternative approaches
may lead to somewhat different parameters, we believe that the
overall trend of two broad relaxations exchanging the role of the
dominating process depending on
temperature will remain the same.\\

\subsection{Freon112a}
\begin{figure}
\includegraphics[width=0.75\textwidth]{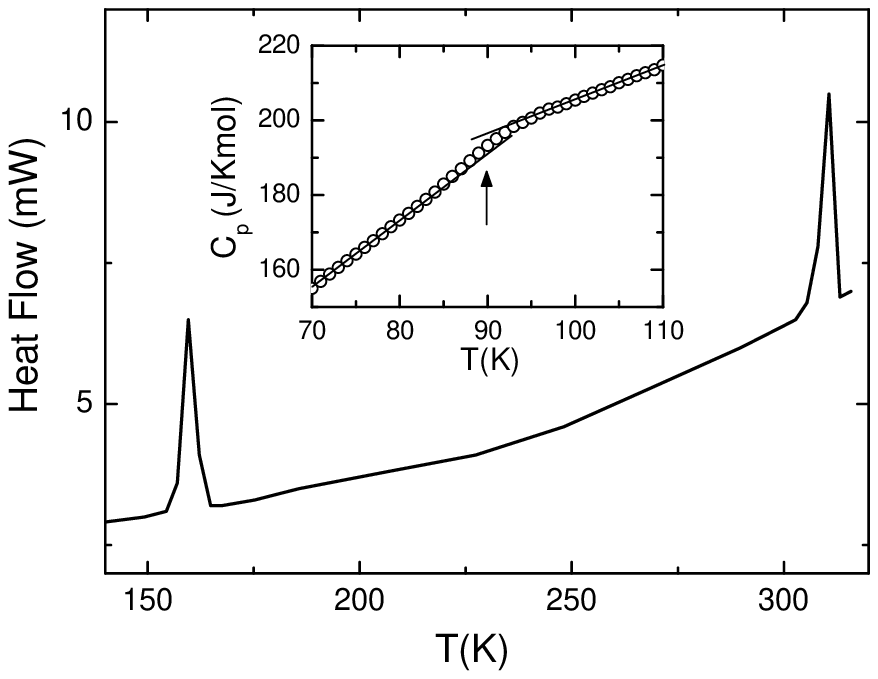}
\caption{\label{CpDSCF11} \footnotesize DSC experiment showing the
polymorphism of Freon112a. The inset shows the specific heat
measurement performed on phase II (see text).}
\end{figure}
In contrast to Freon112, Freon112a (having the same molecular
formula as Freon112) has only one molecular conformation
exhibiting a permanent dipolar moment, which makes worthwhile the
comparison of these two compounds. Since to our knowledge there is
no information on the polymorphism of this substance in the
literature, DSC experiments were performed to establish the phase
sequence for this compound, revealing the existence of two phase
transitions (see Fig. \ref{CpDSCF11}). The obtained melting
temperature ($T_{m}$ = 309 K) and enthalpy change ($\Delta H=3.5
\textrm{ kJmol}^{-1}$) lead to an entropy of fusion of $\Delta
S=11.3 \textrm{ J(K mol)}^{-1}$. Following  Timmermans criterion
($\Delta S_{m} <21 \textrm{ J(K mol)}^{-1}$ for a plastic phase)
\cite{timmermans1961}, this melting entropy value indicates that
the solid phase just below the melting (from now on phase I) can
be assigned as a plastic crystal, similar to the findings in
Freon112. Preliminary neutron diffraction experiments have also
been performed on this compound \cite{neutron}, and the spectra
could be fitted assuming a BCC lattice with a parameter close to
that of the BCC phase of Freon112. However, unlike Freon112,
Freon112a shows a second phase transition from the plastic phase
to a more stable solid phase (from now on phase II) at T = 158 K
($\Delta H=3.4 \textrm{ kJmol}^{-1}$ and $\Delta S=21.6 \textrm{
J(Kmol)}^{-1}$). In Figure \ref{CpDSCF11} specific heat results
obtained with an AC calorimeter in phase II (the only one showing
a glass transition), and DSC experiments used to determine the
polymorphism of the substance, are shown. The figure reveals a
specific heat anomaly at 90$ß\pm$2 K, which can be assigned to a
glass transition, its $T_{g}$ coinciding with that for Freon112.
Thus it seems that some disorder remains in phase II, which
finally freezes at 90 K. The situation is similar to that found in
various other plastic crystals, e.g., some methylbromomethanes
\cite{bromos}, adamantanone \cite{PCdynamics} and ortho-
\cite{lunkenheimer1996} and meta-carborane \cite{PCdynamics}. In
these cases, typical glass phenomena occur in a phase allowing for
restricted orientational motions only, in contrast to the
high-temperature plastic phase where free reorientations of the
molecules are possible \cite{carborane,bromos}. A similar scenario
can be tentatively attributed to Freon112a. As was already
speculated in \cite{kishimoto1978,suga1985}, it seems likely that
in Freon112 the free rotator phase is conserved down to the lowest
temperatures due to the trans-gauche conformational disorder
arising below 130 K.\\

\begin{figure}
\includegraphics[width=0.75\textwidth]{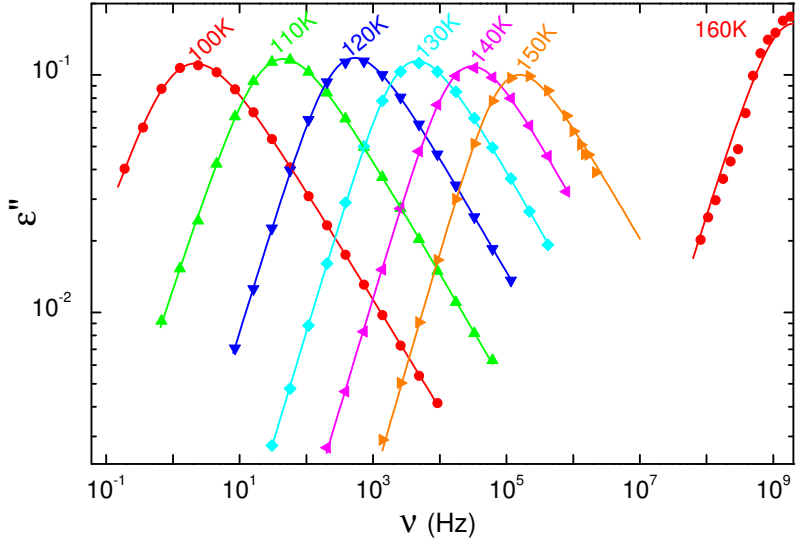}
\caption{\label{efreon11} \footnotesize Dielectric loss for
Freon112. The lines show the result of  fits using the HN
relation. }
\end{figure}

In Fig. \ref{efreon11}, the dielectric loss spectra of Freon112a
are shown. In phase II we observe the typical
temperature-dependent shift of well-pronounced relaxation peaks
over many decades of frequency, proving that indeed glassy
freezing occurs in this phase. Comparing the peak at 160 K, which
was measured in phase I, to that at 150 K in phase II, reveals a
strong slowing down of relaxation at the phase transition. A very
similar behavior was also found in ortho- and meta-carborane and
adamantanone and ascribed to the restriction of the
reorientational motion in the low-temperature phase
\cite{lunkenheimer1996,PCdynamics}. In contrast to the two
successive relaxations found in Freon112 (Fig. \ref{efreon12}), in
Freon112a only one relaxation process seems to prevail. For all
temperatures investigated, the loss peaks shown in Fig.
\ref{efreon11} could well be fitted with a single HN function
(lines). The resulting temperature dependent relaxation time is
shown in Fig. \ref{logtaug}. In marked contrast to the finding in
Freon112,  $\tau(T)$ in phase II of Freon112a follows an Arrhenius
behavior with an energy barrier of 300 meV. Interestingly,
$\tau$(160 K) measured in phase I of Freon112a matches the value
obtained for Freon112 at the same temperature. The remaining
relaxation parameters of Freon112a, showing a smooth and
unspectacular temperature dependence, are given in Fig.
\ref{parameter}.

\section{Discussion}
\begin{figure}
\includegraphics[width=0.75\textwidth]{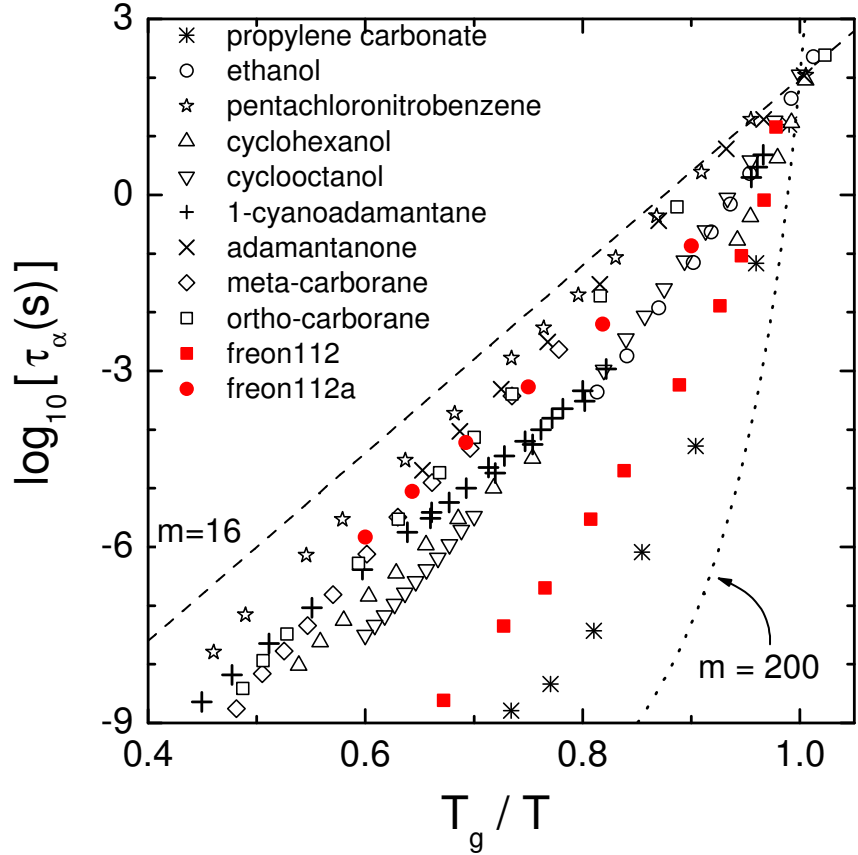}
\caption{\label{angell} \footnotesize Angell plot of the $\alpha$
relaxation times for plastic crystals \cite{PCdynamics}. The
results of the present work are presented as closed symbols. For
comparison results on supercooled-liquid propylene carbonate are
shown. ´The dashed and dotted lines show the behavior for maximal
and minimal fragility.}
\end{figure}

In Fig. \ref{angell}, an Angell-plot \cite{angellplot} of the
temperature dependent $\alpha$-relaxation time for the two
investigated Freon compounds is given, enabling the direct
comparison of the temperature characteristics of $\tau$ to that of
several other plastic crystals \cite{PCdynamics} also shown in the
figure. For comparison, also the curve for a typical fragile
structural glass-former, propylene carbonate, is given
\cite{schneider1999}. As was already predicted by Angell and
coworkers \cite{freonstrong,angell1991} based on a rather
restricted data set, Fig. \ref{angell} demonstrates that Freon112
indeed exhibits pronounced fragile characteristics and in fact it
seems to be the most fragile plastic crystal known so far. As a
quantitative measure of the fragility, the fragility parameter
\textit{m} was introduced in \cite{plazek1991,bohmer1992}, defined
by the slope at $T_{g}$ in the Angell plot. From Fig.
\ref{angell}, we find \textit{m} = 68, characterizing Freon112 as
a fragile glass former. In \cite{PCdynamics} it was speculated
that molecules with high symmetry may exhibit the lowest
fragilities. Considering that many plastic crystals are formed by
rather symmetric molecules, this notion could explain the low
fragility of most members of this group of materials. Freon112,
comprising roughly dumbbell-shaped molecules, certainly is less
symmetric than other plastic crystals, e.g. the nearly spherical
carboranes (\textit{m} = 20 - 21
\cite{lunkenheimer1996,PCdynamics}) or the ring-shaped thiophene
(\textit{m} = 16) \cite{plazek1991,thiophene}. However, the same
could be said for Freon112a, which is revealed as rather strong in
Fig. \ref{angell}. In \cite{PCdynamics}, an explanation for the
low fragility of plastic crystals in line with ideas
\cite{angell1994} that relate the fragility to the form of the
potential energy landscape in configuration space was considered.
The typical properties of fragile and strong glass-formers can be
rationalized assuming that the density of minima in the potential
energy landscape increases with increasing fragility. For plastic
crystals, in \cite{PCdynamics} the lattice symmetry was presumed
to lead to a reduced density of energy minima, which explains
their relatively low fragility. Within this framework the higher
fragility in Freon112 may be rationalized assuming that the strong
trans-gauche conformational disorder in this material leads to an
increased density of energy minima.\\

According to the well-established correlation between fragility
and the width of the  $\alpha$-relaxation peak at $T_{g}$
\cite{bohmer1993,bohmerrelation}, fragile materials should have
rather large $\alpha$-peak widths, corresponding to small values
of the exponent $\beta$, characterizing the high-frequency flank
of the $\alpha$-peak. Indeed, at low temperatures, $\beta_{CD}$ in
Freon112 becomes very small (Fig. \ref{parameter}). With a half
width of about 3.2, the $\alpha$-relaxation in Freon112 even seems
too broad in view of this correlation. However, as due to the
overlap with the  $\beta$-relaxation the determination of the
$\alpha$-peak width at low temperatures has a rather high
uncertainty, it is not possible to arrive at a definite statement
concerning the validity of this correlation in Freon112.

The fragile character of Freon112 is not the only exceptional
feature concerning this compound. Although plastic crystals often
show a $\beta$ relaxation, the dielectric strength of this process
is usually much smaller than that of the $\alpha$ relaxation, and
also smaller than the $\beta$-process found in typical canonical
glass formers \cite{PCdynamics}. That is not the case for
Freon112, as becomes obvious in Fig. \ref{parameter}. At
temperatures above the glass transition, the dielectric strength
of the $\alpha$ relaxation decreases while that of the $\beta$
process increases, the latter even becoming the dominating
relaxation process at high temperatures. This unusual behavior,
namely a $\beta$-process with an amplitude becoming comparable or
even exceeding that of the $\alpha$-process at high temperatures,
to our knowledge so far was never observed in any plastic crystal
and is realized only in few canonical glass formers, e.g. toluene
or polybutadiene \cite{arbe1996,kudlik1999}. In general, the
occurrence of $\beta$-relaxations seems to be a common feature of
canonical glass formers. Sometimes intramolecular motions are held
responsible for $\beta$-relaxations, but Johari and Goldstein
\cite{johari1970} demonstrated that secondary relaxation processes
also show up in relatively simple molecular glass-formers, where
intramolecular contributions seem unlikely. This led to the notion
that these so-called Johari-Goldstein (JG) $\beta$-relaxations may
be inherent to glass-forming materials in general. However, the
microscopic processes behind this kind of $\beta$-relaxations are
still controversially discussed. With an Arrhenius behavior of
$\tau_{\beta}$ below $T_{g}$, the  $\beta$-relaxation in Freon112
detected in the present work exhibits a typical feature of JG
$\beta$-relaxations found also in canonical glass formers
\cite{kudlik1999,johari1970}. Another typical property of JG
relaxations is a correlation of the $\beta$-relaxation time at
$T_{g}$ with the Kohlrausch exponent  $\beta_{KWW}$ promoted in
\cite{ngai1998}, which was rationalized within the framework of
the coupling model \cite{ngai1994}. The fulfillment of this
relation recently even was proposed as criterion to distinguish
genuine JG $\beta$-relaxations (inherent to the glassy state of
matter) from other types of secondary relaxations (e.g., due to
intramolecular modes) \cite{ngai2004}. This relation was
successfully tested for a variety of plastic crystals
\cite{PCdynamics} and for toluene and polybutadiene
\cite{ngai1998}, which show a similar relaxation scenario as
Freon112. However, using the predicted relation $\tau_{\beta} =
t_{c}^{1-\beta_{KWW}}\tau_{\alpha}^{\beta_{KWW}}$ with $t_{c}$ = 2
ps \cite{ngai1998,ngai2004} and the present
$\beta_{KWW}(T_{g})\approx 0.34$, calculated from
$\beta_{CD}\approx$ 0.2 (cf. Fig. \ref{parameter})
\cite{lindsay1980} we obtain $\tau_{\beta}(T_{g})\approx 10^{-7}
s$, which is far-off the actual relaxation time of about 0.1 s
(see Fig. \ref{logtaug}). Thus at first glance it seems that the
observed
$\beta$-relaxation in Freon112 is not a JG relaxation.\\

In contrast to Freon112, Freon112a only shows a single relaxation
process. As one of the main differences of both materials is the
possibility of the first to have two molecular conformations, it
seems likely that the occurrence of two relaxations in Freon112 is
somehow connected to this difference. The transition between the
gauche and trans conformations in Freon112 corresponds to a
rotation around the C-C axis. As this is a dipolar active motion,
one may attribute the second relaxation in Freon112 to this
intramolecular mode, but a corresponding mode of course is also
possible in Freon112a. However, only in Freon112 it leads to a
transition between a conformer with and without dipolar moment,
i.e. after a gauche-trans transition the corresponding molecule no
longer contributes to the  $\alpha$-relaxation, which (in some so
far unknown way) may lead to the unusual relaxational behavior of
Freon112. A more reasonable explanation arises considering the
fact that only  the gauche conformer has a dipolar moment, meaning
that below 130 K (freezing temperature of the conformational
disorder) the dielectric signal arises from a solution of about
50\% of dipolar molecules in an non-polar medium
\cite{proportion}. A recent study of Blochowicz and R\"{o}ssler
\cite{blochowicz2004} of molecules with high dipolar moment in an
only weakly polar substance (50\% 2-picoline in tri-styrene)
revealed dielectric spectra closely resembling the ones obtained
by us: At low temperatures there is an extremely broadened
$\alpha$-relaxation peak and a well-separated weaker
$\beta$-relaxation. For temperatures just above $T_{g}$ there is a
change of the relative dielectric strength of the primary and
secondary relaxation, the latter becoming larger than the first.
Comparing the properties of the $\beta$ relaxation shown in this
paper\cite{blochowicz2004} with that of the secondary relaxation
phenomenon of Freon112, we find that both processes show similar
properties: The relaxation strength of the $\beta$ process is
almost constant for temperatures below $T_{g}$, the time constant
follows an Arrhenius law for $T<T_{g}$ and the relationship
between the activation energy of the $\beta$ process and $T_{g}$
($\Delta H_{a}/(k_{B} T_{g})\approx 25$), promoted in
\cite{kudlik1999,kudlik1998}, is also accomplished taking into
account the aforementioned uncertainty in the determination of
$\Delta H_{a}$ because of non-equilibrium effects for $T<T_{g}$
($\Delta H_{a}/(k_{B} T_{g})= 30$ in our case). Furthermore, if we
check now if the 50\% mixture of 2-picoline in tri-styrene follows
the correlation between the stretching exponent of the
$\alpha$-relaxation and the $\beta$ relaxation time
\cite{ngai1998}, it fails as in the case of Freon112, again
because of the extremely small slope of the high-frequency flank
of the primary relaxation peak. However, in \cite{cappacioli2005}
arguments were put forward that the broadening caused by
concentration fluctuations in the mixture may prevent the width
parameter of the $\alpha$-relaxation being directly determined
from the frequency dependence of the loss. Such a scenario may
also be in effect in the present case of polar Freon112 molecules
with gauche conformation, "dissolved" in an non-polar medium
formed by the trans molecules. As the conformational transitions
in Freon112 exhibit glassy freezing close to 130 K
\cite{kishimoto1978,suga1985}, much higher than $T_{g}\approx 88
K$ determined from $\tau_{\alpha}(T)$ of the $\alpha$-relaxation,
the gauche-trans disorder can be regarded as static on the time
scale of the  $\alpha$-relaxation. Thus there is an effective
substitutional disorder of gauche and trans molecules, which may
be the analogue to the concentration fluctuations in the
picoline/tri-styrene mixture. Therefore the afore-mentioned large
discrepancy of the experimental $\tau_{\beta}(T_{g})$ from the one
calculated using $\beta_{KWW}(T_{g})$ = 0.34, not necessarily
implies the failure of the correlation promoted in \cite{ngai1998}
or that the $\beta$-relaxation in Freon112 is not a JG relaxation
\cite{ngai2004}.\\

Assuming that this  $\beta$-relaxation indeed is of JG type, its
increased strength in the diluted case may be rationalized within
the "islands of mobility" framework promoted, e.g., in
\cite{johari1970}. There it was suggested that JG relaxations in
canonical glass formers arise from localized motions of a fraction
of the molecules residing in regions of lower density, leading to
reduced interactions among neighboring molecules and thus faster
dynamics. It is reasonable that this effect is absent or much
weaker in plastic crystals where density fluctuations are smaller,
being generated by the orientational disorder only, which would
explain the much weaker or even absent  $\beta$-relaxations in
these materials \cite{PCdynamics,brand1999}. The reorientational
dynamics of the molecules in most plastic crystals can be assumed
to be determined by dipolar and steric interactions with
neighbours, which due to the regular crystalline lattice show
smaller spatial fluctuations than in canonical glass formers.
However, the situation is different in conformationally disordered
Freon112, where one could imagine regions of relatively isolated
dipolar gauche molecules in a matrix of nonpolar trans molecules.
These regions could correspond to the less dense regions
considered in \cite{johari1970} for the explanation of the
$\beta$-relaxations in canonical glass formers: Due to the much
reduced dipolar interactions, the reorientational motions in these
regions should be faster, giving rise to the observed strong
$\beta$-relaxation in Freon112.

\section{Conclusion}
In summary, we have performed a thorough dielectric investigation
of plastic crystalline Freon112, comparing the results to those on
Freon112a. In relation to other plastic crystals, Freon112 is
exceptional in many respects: At first, despite it is composed of
rather asymmetric dumbbell-shaped molecules, it forms a plastic
crystalline phase which can easily be supercooled, the transition
to the completely ordered phase being extremely slow
\cite{kishimoto1978,suga1985}. In addition, in contrast to other
plastic crystalline materials \cite{PCdynamics}, the glassy
freezing of the orientational dynamics in Freon112 exhibits
typical fragile characteristics and in fact, with m = 68, to our
knowledge Freon112 is the most fragile orientationally disordered
material known so far. Finally, while other plastic crystals
exhibit no or only weak secondary relaxations
\cite{PCdynamics,brand1999}, Freon112 shows a $\beta$-relaxation
with a dielectric strength comparable to that of the
$\alpha$-relaxation and even becoming the dominant process at high
temperatures. All these exceptional properties are not found in
Freon112a, showing a transition into a phase with restricted
reorientational motions only, strong relaxation characteristics
and neither a  $\beta$-relaxation nor an excess wing. It seems
reasonable to ascribe the unusual findings in Freon112 to the
presence of two molecular conformations in this system, in
contrast to Freon112a, which has only a single one. At first, as
was already considered in \cite{kishimoto1978,suga1985}, the
strong conformational disorder in Freon112, which is frozen on the
time scale of the $\alpha$-relaxation, seems to prevent complete
orientational order at low temperatures as this would imply all
molecules transferring to the same conformation. In addition, the
unusually high fragility may be ascribed to a higher density of
minima in the potential energy landscape caused by the strong
trans-gauche disorder. Finally, this trans-gauche disordered
material may be regarded analoguous to a solution of dipolar
molecules in an non-polar medium, shown to lead to a
$\beta$-relaxation scenario similar to the present observations
\cite{blochowicz2004}, which may be rationalized within the
explantion of the JG relaxation in terms of islands of mobility
\cite{johari1970}. Overall, certainly many of the proposed
explanations of the unusual behavior of Freon112 revealed in the
present work are speculative and need to be corroborated by
further experiments as, e.g., NMR clarifying the details of the
reorientational motions in this compound. However, it seems likely
that Freon112, with its unusual behavior standing out among all
other plastic crystals, may be a key system to enhance our
understanding of glassy dynamics in general and especially of the
nature of the JG process. For this reason currently further
investigations are being carried out in order to determine the
conformational, long- and short- range order in this compound as a
function of temperature by means of neutron diffraction.

\section{Acknowledgements}
We thank C. A. Angell for making us aware of the unusual
properties of Freon112. The authors would  like also to thank
Dr.del Barrio for the DSC measurements in high-pressure cells. One
of the authors (L.C. Pardo) would like to acknowledge the
financial support of the Humboldt Foundation and FIS2005-00975
project from "Ministerio de Educaci\'{o}n y Ciencia".




\begin{thebibliography}{99}
\bibitem{spectra}
The shown curves are restricted to T $\leq$ 159 K as at higher
temperatures the peak frequency lies outside of the investigated
frequency range.
\bibitem{proportion}
This proportion is calculated taking into account that the trans
conformer has about 5-8 meV lower energy than the gauche
conformer, using the relationship $n_{g}/n_{t}$ = 2 exp $(-\Delta
E/RT)$ ($n_{g}$ and $n_{t}$ are the number of gauche and trans
conformers). The factor of two arises  from the fact that there
are two gauche and only one trans conformer.
\end{thebibliography}
\end{document}